\RequirePackage{lineno}
\documentclass[aps,twocolumn,showpacs,amsmath,amssymb,nofootinbib]{revtex4-1}

\usepackage{amsmath}
\usepackage{amsfonts,amssymb}
\usepackage{multirow}
\usepackage[dvips]{graphicx}
\usepackage{float}
\usepackage{appendix}
\usepackage{comment}
\usepackage{color}
\usepackage{url}
\usepackage{overpic}
\usepackage{subfigure}
\usepackage{footnote}
\def\gevcc{\textrm{Ge\kern -0.1em V}/c^2}
\def\gevc{\textrm{Ge\kern -0.1em V}/c}%
\def\gev{\textrm{Ge\kern -0.1em V}}%

\newcommand{\beq}{\begin{equation}}
\newcommand{\eeq}{\end{equation}}
\newcommand{\bea}{\begin{eqnarray}}
\newcommand{\eea}{\end{eqnarray}}
\newcommand {\NIM}    {Nucl.{} Instrum.{} Meth.{} }

\graphicspath{{plots/}}
\RequirePackage{xspace} \allowdisplaybreaks

\begin{document}

%\begin{comment}
%%%%%%%%%%%%%%%%%%%%%%%%%%%%%%%%%%%%%%%%%%%%%%%%%%%%%%%%%%%%%%%
\def\taucharm{$\tau$-charm }
\def \cdots{\cdot\cdot\cdot}
%%%%%%%%%%%%%%%%%%%%%%%%%%%%%%%%%%%%%%%%%%%%%%%%%%%%%%%%%%%%%%%%

\title{A fast simulation package for STCF detector}
\author{%
Xiao-Dong Shi $^{1,2}$\email{xiaodong.shi@mail.ustc.edu.cn}%
\quad Xiao-Rong Zhou $^{1,2}$
\quad Xiao-Shuai Qin $^{3,4}$
\quad Hai-Ping Peng $^{1,2}$
}

\affiliation{
$^1$ State Key Laboratory of Particle Detection and Electronics, Hefei 230026, China\\
$^2$ University of Science and Technology of China, Hefei 230026,  China\\
$^3$ Key Laboratory of Particle Physics and Particle Irradiation, Ministry of Education,Qingdao 266237, China\\
$^4$ Shandong University, Qingdao 266237, China\\
}

%%%%%%%%%
\begin{abstract}
%%%%%%%%%%

A Super Tau Charm Facility (STCF) is one of the major options for the accelerator-based high energy project in China in the post-BEPCII era, and its R\&D program is underway.
The proposed STCF will span center of mass energies ($\sqrt{s}$) ranging from 2 to 7~GeV with a peaking luminosity above $0.5\times 10^{35}$~cm$^{-2}$s$^{-1}$ at $\sqrt{s}=4.0$~GeV, and will provide a unique platform for tau-charm physics and hadron physics.
In order to evaluate the physical potential capabilities and optimize the detector design, a fast simulation package has been developed.
This package takes as inputs the response of physical objects in each sub-system of the detector including resolution,
efficiency as well as related variables for the kinematic fit and the secondary vertex reconstruction algorithm.
%Validated well on both sub-detector responses and physical programs,
It can flexibly adjust the responses of each sub-detector system
and is a critical tool for the STCF R\&D program.

\end{abstract}

\maketitle

%\tableofcontents

%%%%%%%%%%%%
\section{Introduction}
\label{INTRO}
%%%%%%%%%%%%%%
The Standard Model (SM) has successfully explained almost all experimental results in the subatomic world, and is generally accepted as the fundamental theory of elementary particle physics.
Despite its success, the SM leaves many questions unanswered~\cite{newphy}.
Quantum chromodynamics (QCD), the component of the SM that describes the strong interactions between quarks and gluons, exhibits two main properties: color confinement~\cite{confinement} and asymptotic freedom~\cite{asymptotic1,asymptotic2}.
In the high energy region, where the property of asymptotic freedom property facilitates the applications of perturbation theory, QCD has been verified by many different experiments at the level of a few percent. 
However, this is not the case in the low energy region, where quantitative tests of QCD are fewer, mainly due to the need for non-perturbative calculation techniques. 
Here, more sophisticated theoretical and experimental approaches are needed.
Hadrons, which are composite particles made of two or more quarks bound by the strong force, provide an
excellent platform to explore non-perturbative QCD.
In addition, search for new physics beyond the SM is a major mission of particle physics that requires 
mutually complementary efforts covering many different energy regions. 

In this research program, a tau charm facility (TCF), an electron-positron collider operating at the transition energy region between perturbative and non-perturbative QCD, would play a unique role and would be of great interest to the elementary particle physics community.
A TCF has several unique and advanced features: copious production of light hadron resonances, charmonium, and charmed hadrons; 
access to exotic hadrons, gluonic particles, and hybrids; 
threshold production of $\tau$ lepton and charmed hadron pairs $etc$.
In comparison to B factory (Belle II)~\cite{Adachi:2018qme} and hadron collider (LHCb) experiments~\cite{Alves:2008zz}, which also  produce $\tau$ lepton and charmed hadrons copiously, a TCF is at a disadvantage in terms of the absolute number of events.
However, it has an excellent signal to background ratio, high detection efficiency, well controlled systematic uncertainties, capabilities for fully reconstructed event, and it provides clear signals for undetected particles such as neutrinos.

Historically, there have several TCF experiments, such as MARKI-III~\cite{Galison:1992us,Abrams:1989cm,Bernstein:1983wk}, DM2~\cite{Augustin:1980ad} $etc$. 
which had produced remarkable results in testing the SM and searching for physics beyond the SM.
For the TCF experiments in China, the details  will be discussed in next section.

\section {Tau Charm facility in China}
The Beijing Electron Positron Collider (BEPC) and its associated Beijing Spectrometer (BES)~\cite{Bai:1994zm}, located in Beijing, China, is the latest in the sequence of TCFs.
BEPC/BES, which started near the end of 1980s, has produced a number of important physics results, including precision measurement of the $\tau$-lepton mass, $e^+e^-$ annihilation cross section, and the discovery of the X(1835), ~\cite{Fang:2017znv,Ablikim:2005um}.
In 2008, this was upgraded to BEPCII, a double ring collider that covers the center-of-mass energy (CME) between 2.0 and 4.9 GeV, with an unprecedented peak luminosity of $10^{33}$~cm$^{-2}$s$^{-1}$ at $\sqrt{s}= 3.78$ GeV and BESIII, a state of the art detector.~\cite{bepcii,Ablikim:2009aa}
As the only running TCF in the world , BEPCII/BESIII has been very successful and fruitful~\cite{Ablikim:2019hff}.
After more than ten years operation, BEPCII has surpassed its designed luminosity and extended its CME up to 4.9~GeV.
Recently, BESIII finished collecting 10B $J/\psi$ events for the study of light hadron physics, thereby completing one of its long-term opeartion goals.
It has also collected a 2.9~fb$^{-1}$ data sample of $\psi(3770)$ decays for studies on charmed meson, $\sim$0.7~fb$^{-1}$ data with CME between 2.0 and 3.0 GeV for precision measurements of the total cross section for $e^+e^-$ annilhilation into hadrons ($\sim$3\% uncertainty),
and other low energy QCD studies, and a $\sim$20~fb$^{-1}$ data sample with CME between 4.0 and 4.7 GeV for charmonium-like XYZ, charmed baryon ($\Lambda^+_c$) and strange-charmed meson ($D_{s}$) studies.
With these data samples, BESIII has published the discovery of $Z_{c}(3900)$, the observation of the large isospin violation in the $\eta(1405)\to f_0(980) \pi^0$ process, the abrupt change in the $X(1835)$ line shape at the $p\bar{p}$ threshold, precision measurements of the $D$ leptonic decays, and the production cross section of $e^+e^-\to \pi^+\pi^-$, the precision measurement of strong phase differences in $D$ decays, the observation of $Z_{cs}(3985)^-$,  as well as the precision  measurements of $\Lambda^+_c$ decays~\cite{Ablikim:2013mio,BESIII:2012aa,Ablikim:2016itz,Ablikim:2016sqt,Ablikim:2015orh,Ablikim:2015flg,Ablikim:2020yif,Ablikim:2020hsk}.

Despite its success, with the deepening of research and a comprehensive  understanding of the micro world, the physics potential of BEPCII/BESIII is limited by its luminosity and CME. For example, the understanding of the internal composition of XYZ particles and their underlying dynamics requires larger luminosity and extended CME, the study of charmed baryons physics requires extended CME, studies of charmed mesons and $\tau$ physics requires more luminosity.
Furthermore, as we know, the Belle II experiment is under commissioning, and is expected to accumulate 50~ab$^{-1}$ data by year 2024~\cite{Kou:2018nap}; the LHCb is on the upgrade, and is expected to collect much more data in future~\cite{Bediaga:2018lhg}.
Both Belle II and LHCb experiments are challenging BESIII in physics potential, but they are also requiring more precise inputs from the TCF, $e.g.$, in the measurement of the strong phase difference in charmed meson decay for the precision measurement of the Cabibbo-Kabayashi-Maskawa (CKM) element $\gamma/\phi_3$.
Limited by the space of storage ring tunnel, BEPCII has no potential for a major upgrade with new accelerator designs, such as nano-beam, large Piwinski angle, and crab waist techniques.~\cite{Akai:2018mbz}
Therefore, a super tau charm facility (STCF) is proposed as a natural extension and a viable option for an accelerator based high energy project in China in the post BEPCII/BESIII era.

\section{The new generation STCF in China}

The proposed STCF~\cite{STCF} in China is a symmetric double ring electron-positron collider.
It is designed to have CME ranging from 2 to 7~GeV, 
and peaking luminosity beyond $0.5\times 10^{35}$~cm$^{-2}$s$^{-1}$ at $\sqrt{s}=4$~GeV (2 orders higher than BEPCII's). 
It also leaves the potential for upgrading to higher luminosity and implementing polarized  beams in the future~\cite{Luo:2019gri}.
To achieve such high luminosity, several advanced technologies, such as the crabbed waist and large Piwinski angle collision are implemented at the interaction region~\cite{Ohmi:2017cwi}.
With such high luminosity, STCF is expected to deliver more than 1~ab$^{-1}$ of integrated luminosity per year, which provides an excellent opportunity for a broad range of physics studies at the tau-charm energy region including QCD confinement, hadron physics, flavor physics and CP violation as well as direct searches for new physics.

\begin{figure}[htbp]
\begin{center}
 \includegraphics[width=6.5cm,height=9.2cm,angle=-90]{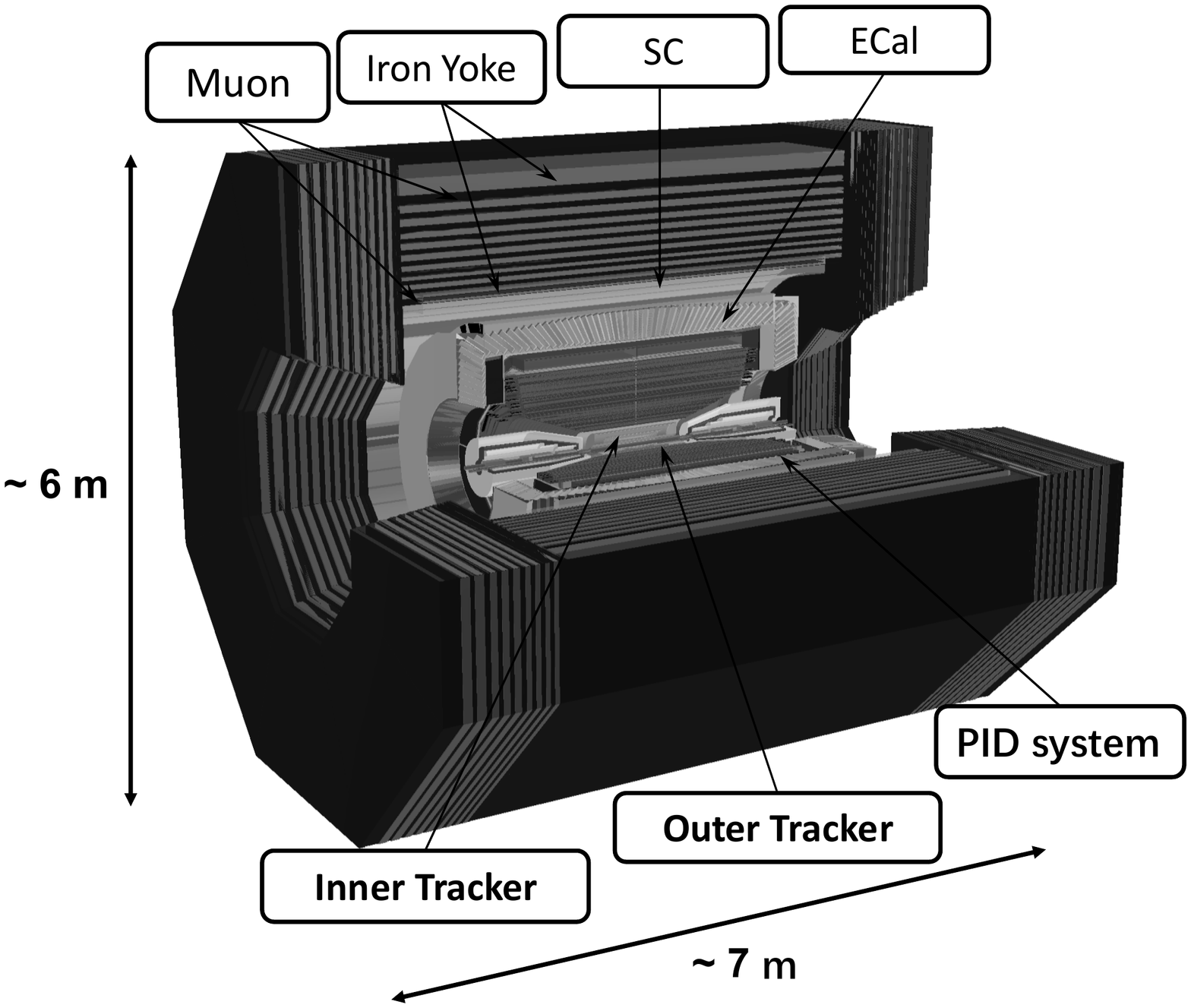} \caption{
   \label{stcf}
    The concept design of the STCF detector, visualized by DD4hep.
}
\end{center}
\end{figure}

At present,  the STCF is at the conceptual design stage.  The detector is designed to maximize the physics potential at the tau-charm energy region and be compatible with the experimental conditions of high luminosity.  
The STCF detector features large solid angle coverage, low noise, high detection efficiency and resolution and excellent particle identification capability. 
It is also required to be of high rate capability and able to operate efficiently under high level of radiation background.
From the interaction point outwards, the STCF detector consists of an inner tracker, an outer traker, a particle identification (PID) system, an electromagnetic calorimeter (EMC), a super-conducting solenoid and a muon detector (MUD). 
Among all the sub-detectors, the inner tracker is the closest one to the interaction point, and hence exposed to the highest level of radiation.~\cite{Lewis:2018ayu,Dong:2015kba}
To tolerate the ultra-high radiation background,
a novel micro-pattern gaseous detector, based on the uRWELL technology and consisting of three cylindrical layers located at 6, 11 and 16 cm away from the interaction point, is proposed to be a baseline option for the inner tracker.  
A helium-gas-based cylindrical main drift chamber (MDC), spanning from 200 to 820~mm in radius, is proposed to be the outer tracker.  The momentum resolution in a 1 T magnetic field is expected to be better than 0.5\% for charged tracks with a momentum of 1~$\gevc$. The dE/dx resolution is better than 6\%, which can be exploited to serve the particle identification for low momentum charged particles.
The PID system uses a Ring Imaging Cherenkov (RICH) detector in the barrel region and a Detection of Internally Reflected Cherenkov (DIRC) detector in the endcap regions to achieve a 3$\sigma$ separation between kaons and pions with a momentum up to 2~$\gevc$.   Separation capability between muons and pions of 3$\sigma$ is also possible with a momentum between 0.2 and 0.6~$\gevc$.
A crystal calorimeter based on pure CsI crystals read out with APDs for energy measurement and SiPMs for precise timing, is proposed for the EMC to achieve an excellent energy resolution (better than 2.5\% with an energy 1~$\gev$) and a good time resolution ($\sim$300 picoseconds for photons) in a high radiation background.  The timing capability of the EMC allows to effectively separate photons from neutrons and $K_{L}^{0}$ in the energy region of interest. 
To achieve a higher spatial resolution, the crystal size will be optimized.
A super-conducting solenoid magnet surrounding the EMC provides the tracking system with a magnetic field of 1 T.
A hybrid of multi-gap resistive plate chamber (3 inner layers) and plastic scintillator (7 outer layers) detector is proposed as the baseline option for the MUD,
and provides excellent capability to efficiently separate muons from pions with a mis-identification rate less than 3\%. 
The conceptual design of the STCF detector is shown in Fig.~\ref{stcf} and is fully described by DD4hep ~\cite{Petric:2017psf}.

\section{Fast Simulation of the detector response}
%%%%%%%%%%%%%%
\label{sec2}

To investigate the physics capability and to optimize the detector design, a fast simulation package dedicated to the STCF detector has been developed.
The package is developed based on the BESIII Offline Software System (BOSS)~\cite{BOSS}, which will minimize the amount of work for code development, improve the reliability and stability of package,
and make it friendly to the users, especially for those with data analysis experience on the BESIII experiment.
%To implement the analysis tools developed by BESIII, including vertex finder, kinematic fit and second vertex fit packages {\it etc}, as well as to perform the analysis in the BOSS framework, the data structure of physical observables used in the analysis is inherited from the BESIII experiments, and can be extended if any new variable is available by the refined STCF detector but absent in the BESIII experiment, and willing to be used in the physics analysis.
The data structure of physical observables is inherited from BESIII experiment, 
so the analysis tools developed by BESIII can be implemented, including vertex finder, kinematic fit and second vertex fit packages {\it etc}. 
It can be extended if any new variable is available by the refined STCF detector but absent in the BESIII experiment.
Instead of performing the detailed simulation of every interaction of the final objects in each sub-detector by {\sc Geant4},
we model the responses of objects in each sub-detector, including efficiency, resolutions and others variables used in data analysis, by sampling the shape and magnitude
according to their performance.
%we model the responses of objects in each sub-detector, including efficiency, resolutions (spatial, momentum, energy, time {\it etc}) and others variables used in data analysis, by sampling randomly according to the shape and magnitude of their performance.
The performances for a given type particle in each sub-detector are described by the empirical formulas or extracted from simulation of the BESIII/STCF detector. 
A brief description of the modeling of responses follows below.
By default, all the parameterized parameters for each sub-detector performance are set based on the BESIII performance~\cite{Ablikim:2009aa}, but can be adjusted flexibly by a scale factor according to the expected performance of the STCF detector, or by implementing a special interface to model any performance described with an external histogram, an input curve, or a series of discrete data.
All of these configuration are interfaced conveniently. 

In this package, the modeling of the responses for each particle type are implemented individually.
The detection efficiency of each sub-detector is simulated by a sampling including the dependence on the particle type, momentum, and $\cos\theta$, where $\theta$ is the polar angle of objects in the laboratory frame.
To simulate the observable of each sub-detector, such as energy, momentum, spatial and time $etc$, 
the MC truth value is smeared with the resolution sampled by the corresponding distribution.
%To simulate the observable of each sub-detector, such as energy, momentum, spatial and time $etc$, the corresponding resolution is sampled according to its distribution and overlaying on the MC truth value.
%The detector efficiency is simulated by a sampling according to its curves as a function of the two-dimension variables, $i.e.$ momentum versus $\cos\theta$, where $\theta$ is the polar angle of objects in the laboratory frame.
%For the observables with measurement uncertainty, such as energy, momentum, spatial and time $etc$, the simulated value is the overlaying of the detection resolution on top of the MC truth value, where the corresponding detection resolution is extracted by sampling according to their distributions.

The MDC system handles the reconstruction of all charged particles. 
In this package, the tracking efficiency, the spatial and momentum resolution of the MDC are simulated. 
The error matrix from the helix fit, which will be used in the analysis for the kinematic fit is also simulated.
The EMC system handles the measurement of energy deposition for photons and electrons, where the detection efficiency, both the energy and spatial resolution, as well as their uncertainties are implemented.
The responses of other charged particles ($\pi^{\pm}$, $K^{\pm}$, $\mu^{\pm}$ and $p(\bar{p})$), and neutral particles ($K_{L}^0$ and neutron) in the EMC, are also taken into consideration.
However, since the responses of these particles are not sensitive to the design of EMC, they are fixed to those extracted from BESIII detection.
Additionally the variables for the separation of photons from neutrons, the shower shapes and the timing measurement, are also simulated.
For the STCF detector, the $dE/dx$ from the MDC and the information from the RICH and DIRC are used to separate the hadrons $K/\pi/p$.
To simplify the simulation and reconstruction processes, instead of simulating the details of each PID detecter's response, we directly simulate probabilities of the detection efficiency and fake rates between the different hadrons.
For the $\pi/\mu$ separation, the information from the muon detector can also be included. Again, only the probabilities of separation and mis-identification between $\pi$ and $\mu$ are simulated.
For the particles with long lifetime, $i.e.$ $K_{S}^0$ and $\Lambda$, the flight length is one of the most important observables.
However there is no need for special treatment for $\pi/p$ from $K_{S}^0$ or $\Lambda$ except recording the decay point from the generator smeared with a certain resolution.
It is worth to mention that in the fast simulation package, we try to model the responses (observations) in each sub-detector as detailed and realistical as possible. However, the correlations between the observations in the different sub-detectors, even in the same sub-detector (for example, the correlations among non-diagonal parameters in the helix error matrix) are not taken into account.

%%%%%%%%%%%%%%%%%
\section{Package Validation with BESIII Detector}
%%%%%%%%%%%%%
\label{sec3}

\begin{figure}[htbp]
\begin{center}
\begin{overpic}[width=0.19\paperwidth,angle=0]{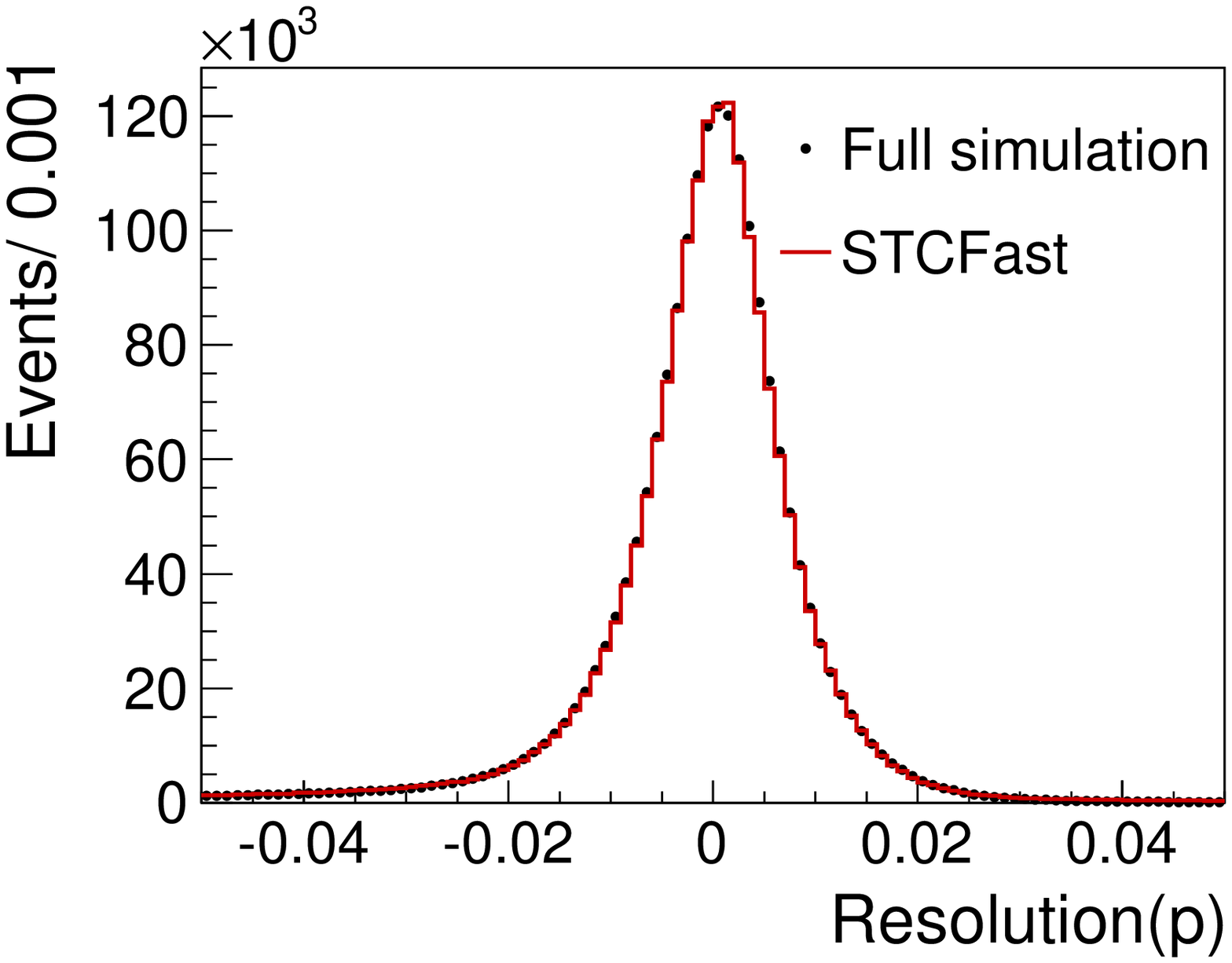}
\put(26,58){\small{(a)}}
\end{overpic}
\begin{overpic}[width=0.19\paperwidth,angle=0]{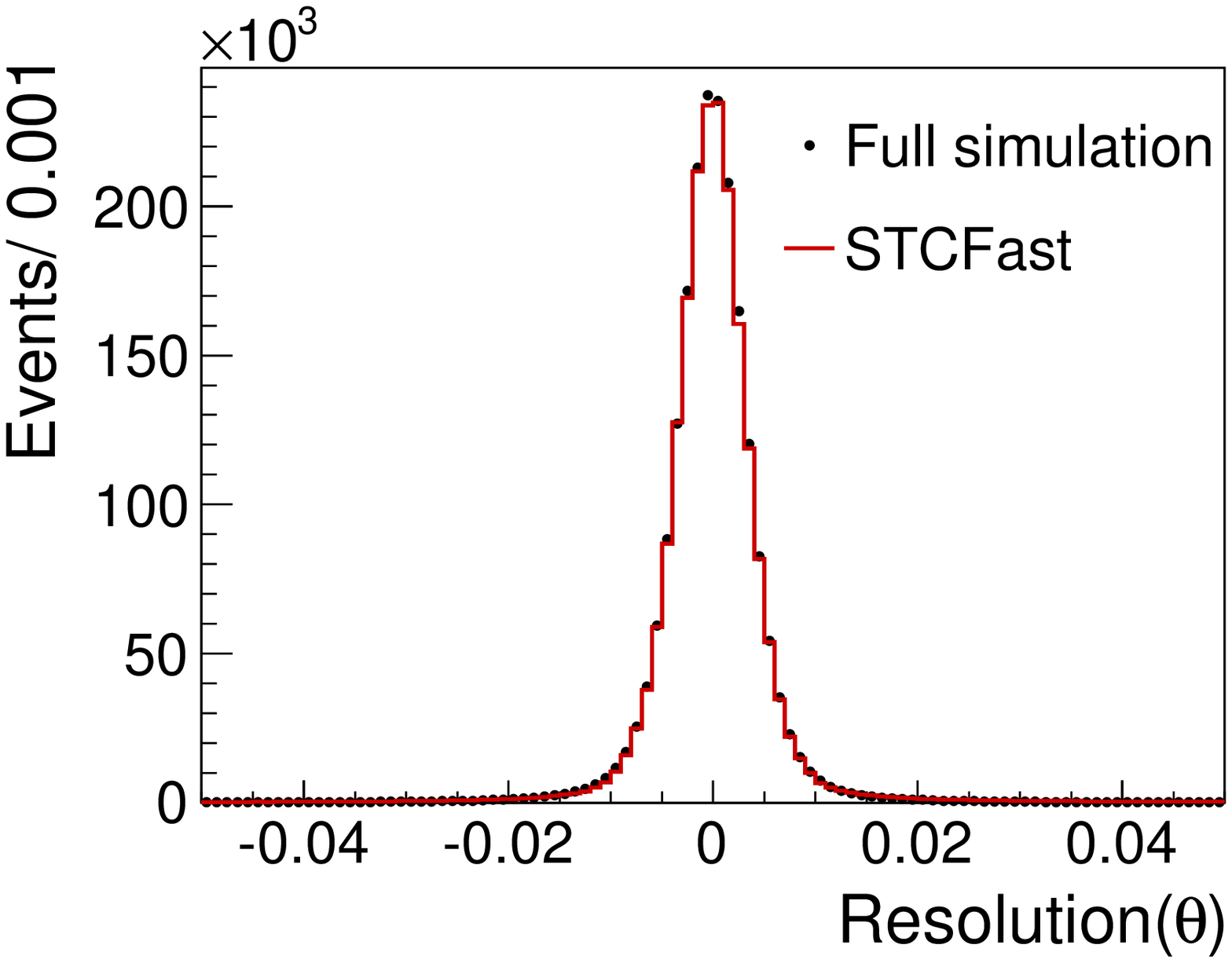}
\put(26,58){\small{(b)}}
\end{overpic}
\begin{overpic}[width=0.19\paperwidth,angle=0]{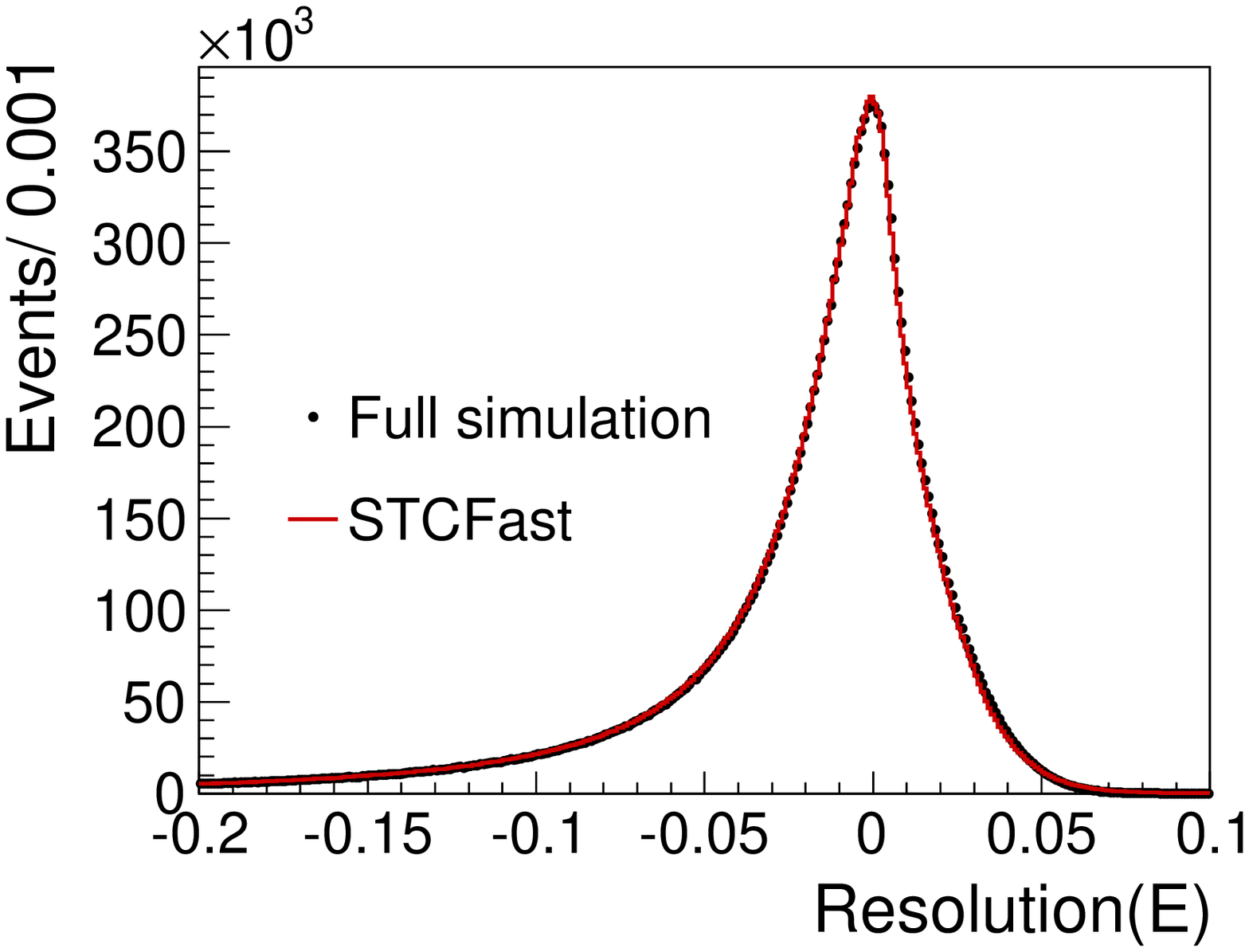}
\put(26,58){\small{(c)}}
\end{overpic}
\begin{overpic}[width=0.19\paperwidth,angle=0]{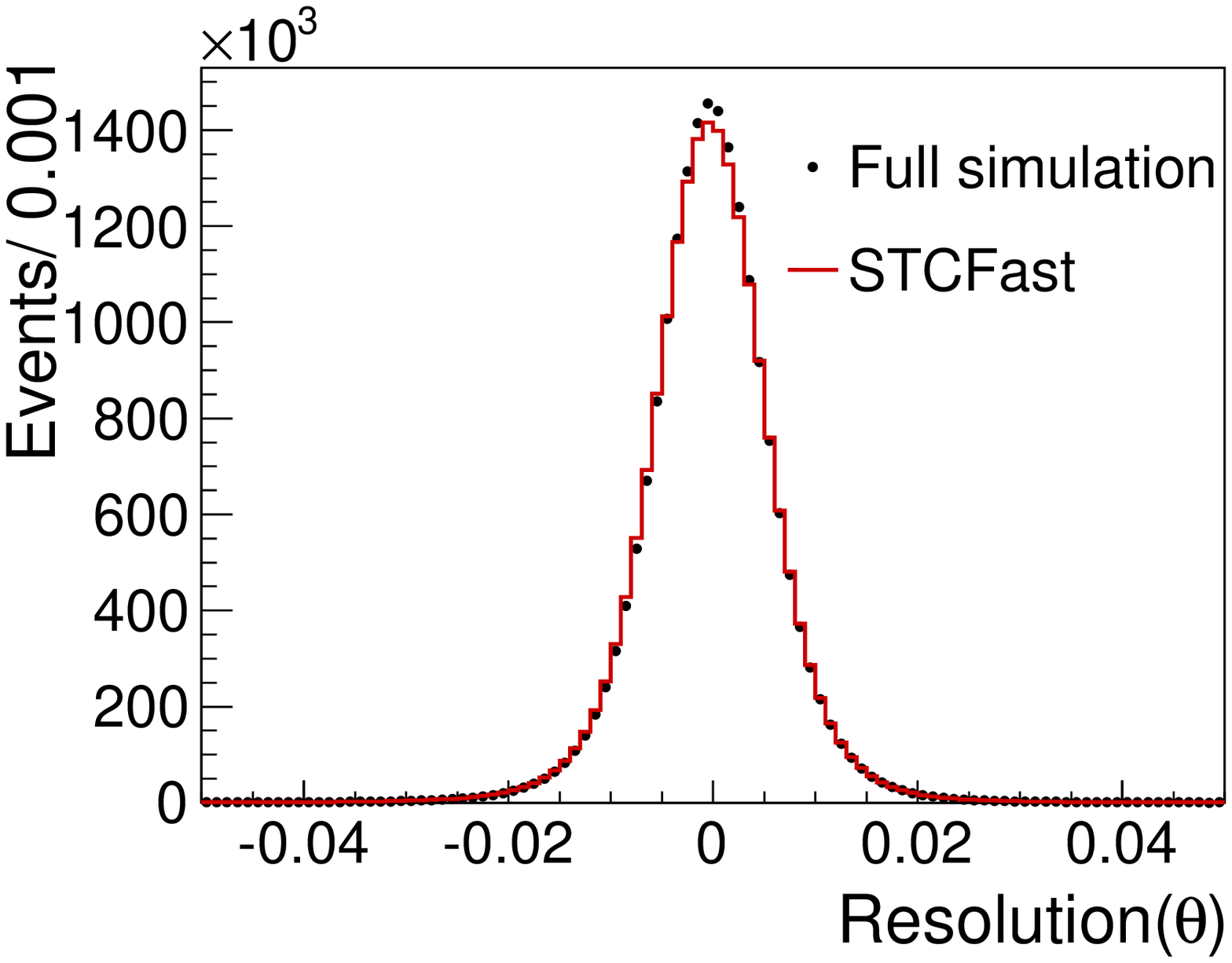}
\put(26,58){\small{(d)}}
\end{overpic}
\end{center}
\vskip -0.6cm
\caption{    Momentum resolution (a) and direction resolution (b) for $e^{\pm}$;
						 Energy resolution (c) and position resolution (d) for $\gamma$.
                Dots with error bars represent full simulation.
                The red line represents fast simulation.}
    \label{pic:obj}
\end{figure}

To check the reliability and stability of the fast simulation package, the validations are performed in different aspects.
At the object level, we have done the checks on all the observations for all particles, and all observations are consistent well within the expectations from the input object. 
We also set the detector performance in the fast simulation package to be those of BESIII detector, and compare the corresponding outputs, including shapes and magnitudes, to those from the BESIII sub-detector responses~\cite{Bai:1994zm}.
%full simulation by {\sc Geant4}. 
Some validations are shown in Figs.~\ref{pic:obj}, where good agreements are achieved. 
In additional, we performed the full physics analysis for some key or interesting processes 
in the fast simulation package by setting the same performance as BESIII detector, and comparing with the results from BESIII's full simulation, {\it e.g.} event selection efficiency, distribution of some physical variables with interest {\it etc}.
The details are given below.

%%%%%%%%%%%%%%%%
\subsection{$e^+e^-\rightarrow\pi^{+}\pi^{-}\psi(3686)$ at $\sqrt{s}=4.26$~GeV}
%%%%%%%%%%%%%

\begin{figure}[htbp]
\begin{center}
\begin{overpic}[width=4.2cm,angle=0]{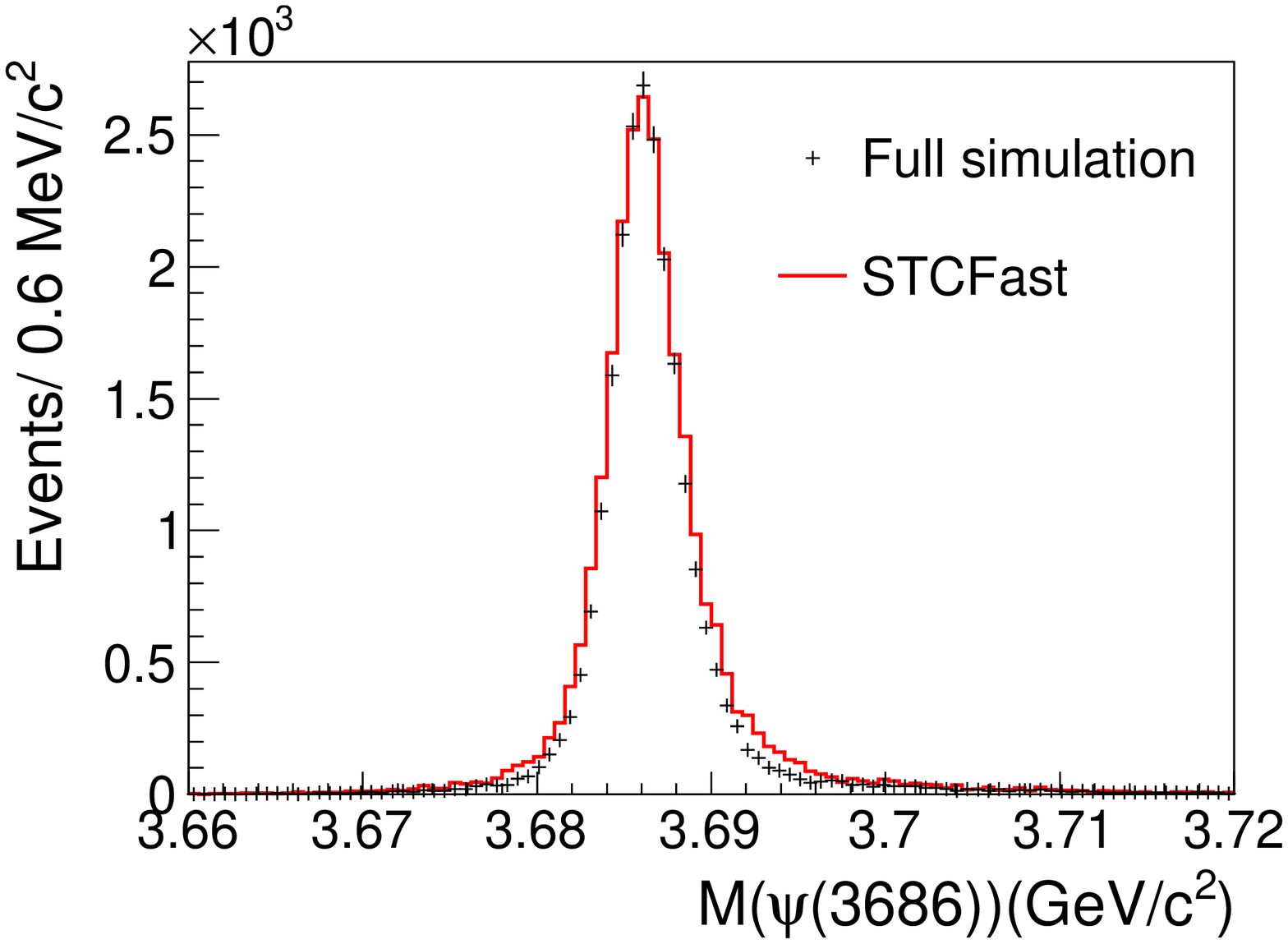}
\put(24,58){\small{(a)}}
\end{overpic}
\begin{overpic}[width=4.2cm,angle=0]{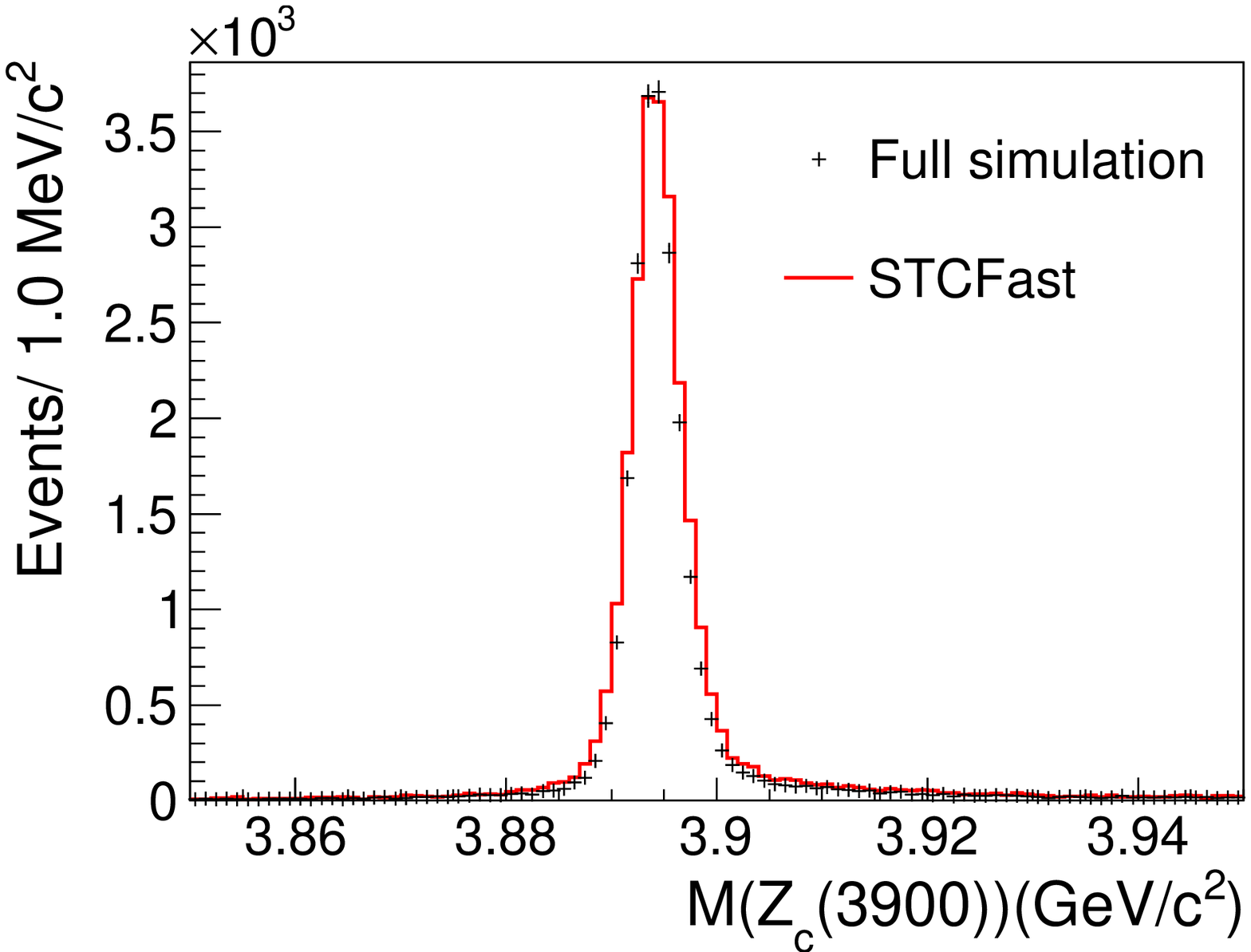}
\put(24,58){\small{(b)}}
\end{overpic}
\end{center}
\vskip -0.6cm
\caption{Invariant mass of (a) $\psi(3686)$ and (b) $Z_c(3900)$ in
		$e^+e^-\rightarrow\pi Z_{c}$, $Z_{c}\rightarrow\pi\psi(3686)$.
		Dots with error bars represent full simulation events.
		The red line represents fast simulation events.}
    \label{zc}
\end{figure}

The decay $e^{+}e^{-}\to\pi^{+}\pi^{-}\psi(3686)$ with subsequent decay $\psi(3686)\to \pi^{+}\pi^{-}J/\psi$ and $J/\psi \to \ell^+\ell^-$ is one of the unique processes to study the charmonium-like $Y$ states and charged $Z_{c}$ states.
Since the process includes four charged pions with low momentum in the final states, it is an ideal process to study the performance of low momentum charged track reconstruction and guide the design of the MDC.
We perform the same event selections as Ref.~\cite{Ablikim:2017oaf}, for both BESIII full MC simulation sample and fast simulation sample and obtain similar event selection efficiencies: 24.9$\pm$0.1$\%$ and 27.7$\pm$0.1$\%$.
To check the impact  of momentum resolution of low momentum charged tracks, we assume that $Z_{c}(3900)$ is an intermediate state in the process $e^{+}e^{-}\to\pi^{+}\pi^{-}\psi(3686)$, and decays to the $\pi^{\pm} \psi(3686)$ final state.
Figs.~\ref{zc} show the comparison of the invariant mass of $\psi(3686)$ and $Z_{c}(3900)$ between the full simulation sample and fast simulation sample. 
Good agreement is achieved in modeling both the core and the tails of the distribution.

Similar tests have been performed on the processes $e^{+}e^{-}\to\phi\pi^{+}\pi^{-}$ at $\sqrt{s}=2.125$~GeV
and $e^{+}e^{-}\to K^{+}K^{-}J/\psi$ at $\sqrt{s}=4.26$~GeV.
%consistencies on mass resolution of intermediate states and the event selection efficiency between full simulation and fast simulation are achieved.
Consistency on the event selection efficiency and mass resolution of intermediate states between full simulation and fast simulation are observed.

%%%%%%%%%%%%%%%%%
\subsection{$J/\psi\rightarrow\gamma \pi^{0}\pi^{0}$}
%%%%%%%%%%%%%

%Glueball hunting is one of key topics in the tau-charm facility.
Searching of exotic states, like glueballs, hybrids, etc., are essential for understanding color confinement.
The $J/\psi$ radiative decay is a gluon-rich environment and is an ideal platform to search for glueballs.
The scalar meson, $f_{0}(1710)$, is produced with large branching fraction (in order $10^{-3}$) in the $J/\psi$ radiative decay, and is regarded as one of the most promising glueball candidates, or a state with a large gluenic component.
The decay of $f_0(1710)$ is useful to investigate its constitution, and the prominent decay $f_{0}(1710)\to\pi^0\pi^0$ is of great interest~\cite{gampi0pi0}.
The cascade decay $J/\psi\to\gamma f_0(1710)\to\gamma \pi^0\pi^0$ consists of purely neutral particles in the final state. 
Consequently it is sensitive to the photon detection and may guide the EMC design.

\begin{figure}[htbp]
\begin{center}
\begin{overpic}[width=4.2cm,angle=0]{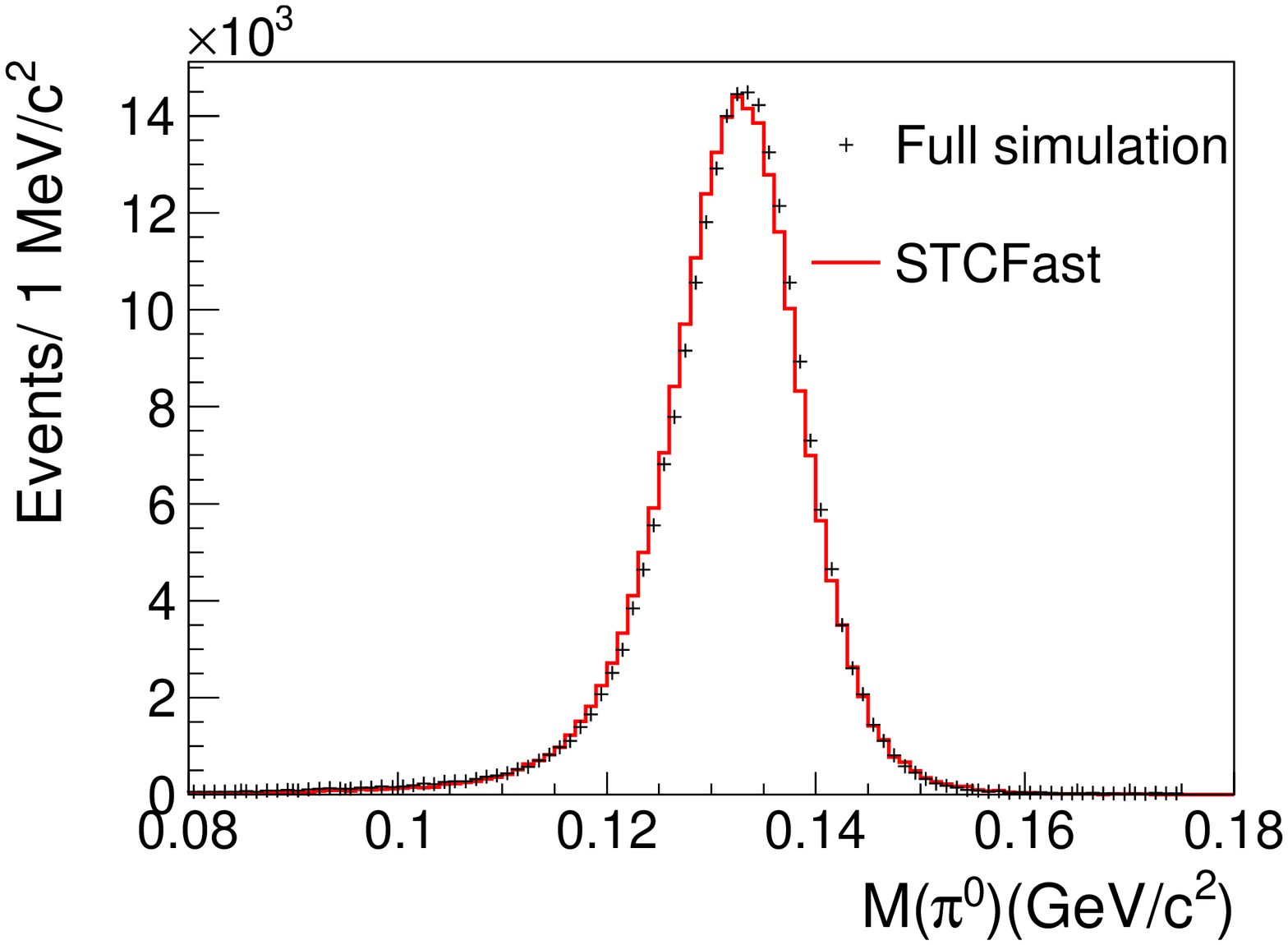}
\put(24,58){\small{(a)}}
\end{overpic}
\begin{overpic}[width=4.2cm,angle=0]{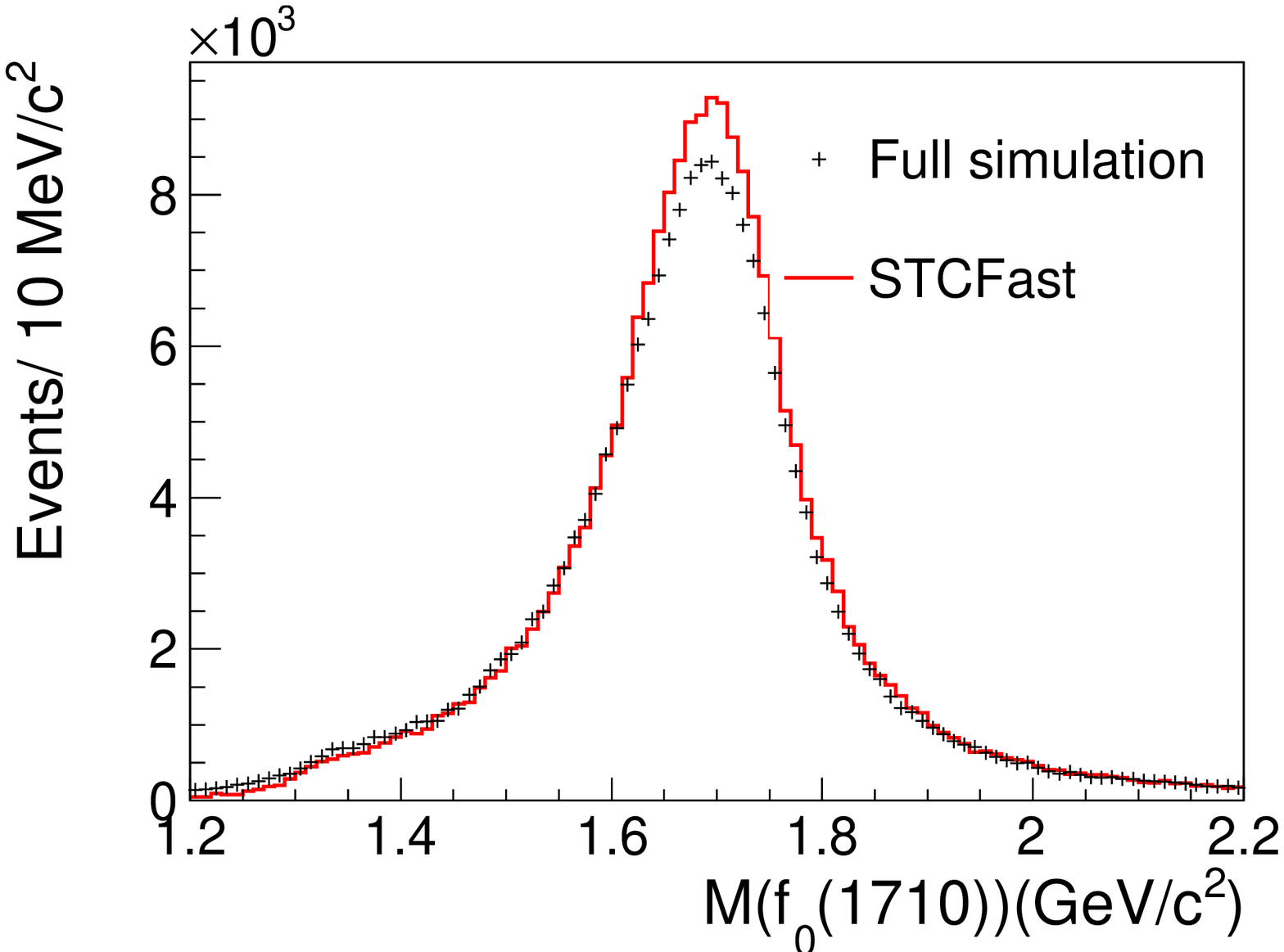}
\put(24,58){\small{(b)}}
\end{overpic}
\end{center}
\vskip -0.6cm
\caption{ Invariant mass of (a) $\pi^{0}$ and (b) $f_{0}(1710)$ in
		$J/\psi\rightarrow\gamma f_0(1710), f_0(1710)\rightarrow\pi^0\pi^0$.
		Dots with error bars represent full simulation events.
		The red line represents fast simulation events.}
    \label{gf0}
\end{figure}

The samples of decay $J/\psi\to\gamma f_{0}(1710)$, $f_{0}(1710)\to\pi^{0}\pi^{0}$
are generated with the BESIII full simulation package and fast simulation package, separately. 
The best photons pairing to a $\pi^{0}$ are selected by
choosing the combination that minimizes $(M_{\gamma_{1}\gamma_{2}}-m_{\pi^{0}})^{2}
+(M_{\gamma_{3}\gamma_{4}}-m_{\pi^{0}})^{2}$, where $m_{\pi^{0}}$ is the mass of the $\pi^{0}$ given by the PDG~\cite{PDG}.
The event selection efficiencies were validated and shown to be consistent between full and fast simulation. 
Figs.~\ref{gf0} show the comparison of the invariant mass of $\pi^{0}$ and $f_{0}(1710)$ candidates between
fast simulation and full simulation. 
Good agreement is achieved for both, which indicates the well modelling of the EMC's performance.

%%%%%%%%%%%%%%%%%
\subsection{Vertex and Kinematic fit}
%%%%%%%%%%%%%

\begin{figure}[htbp]
\begin{center}
\begin{overpic}[width=0.19\paperwidth,angle=0]{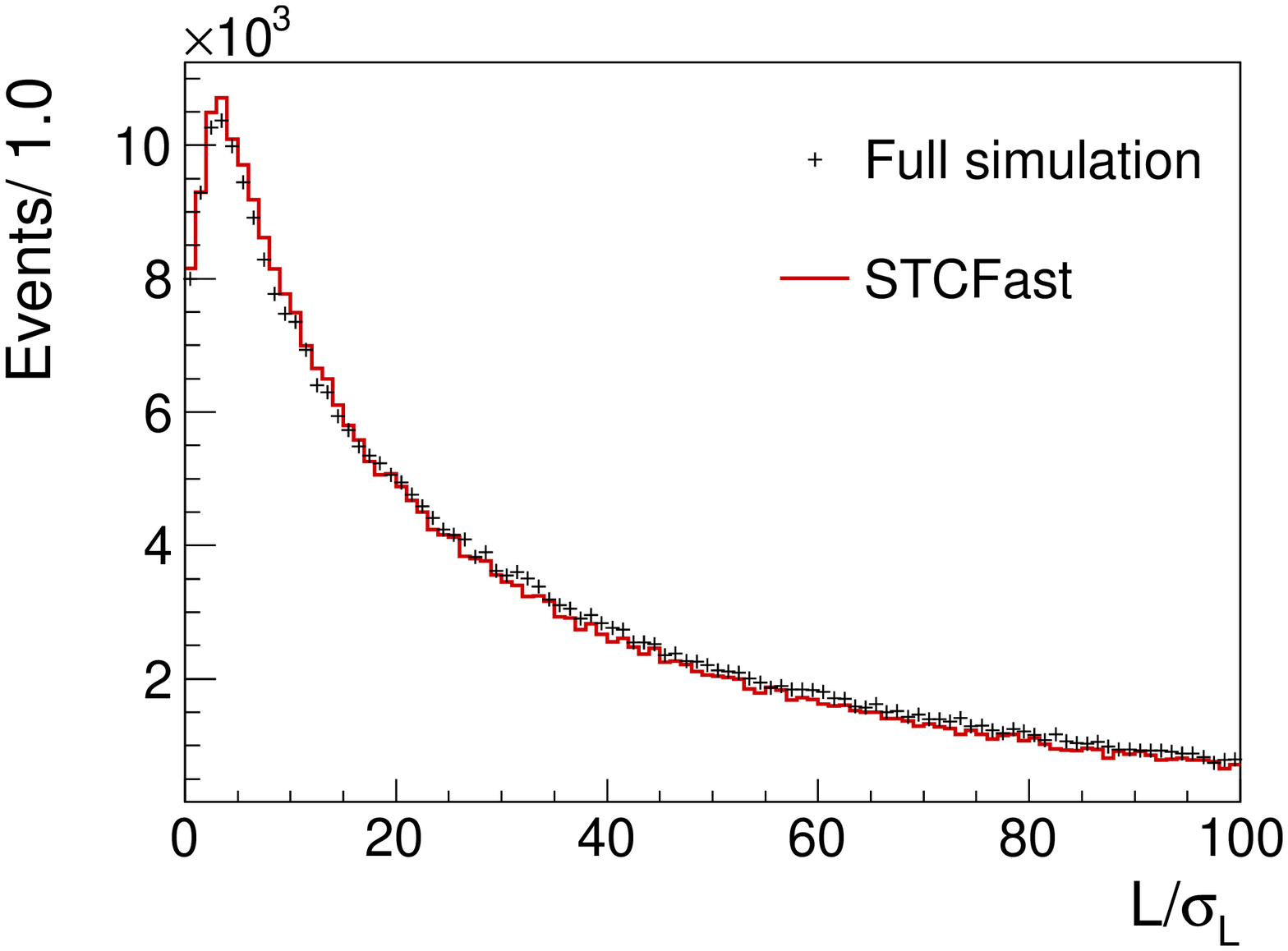}
\put(26,58){\small{(a)}}
\end{overpic}
\begin{overpic}[width=0.19\paperwidth,angle=0]{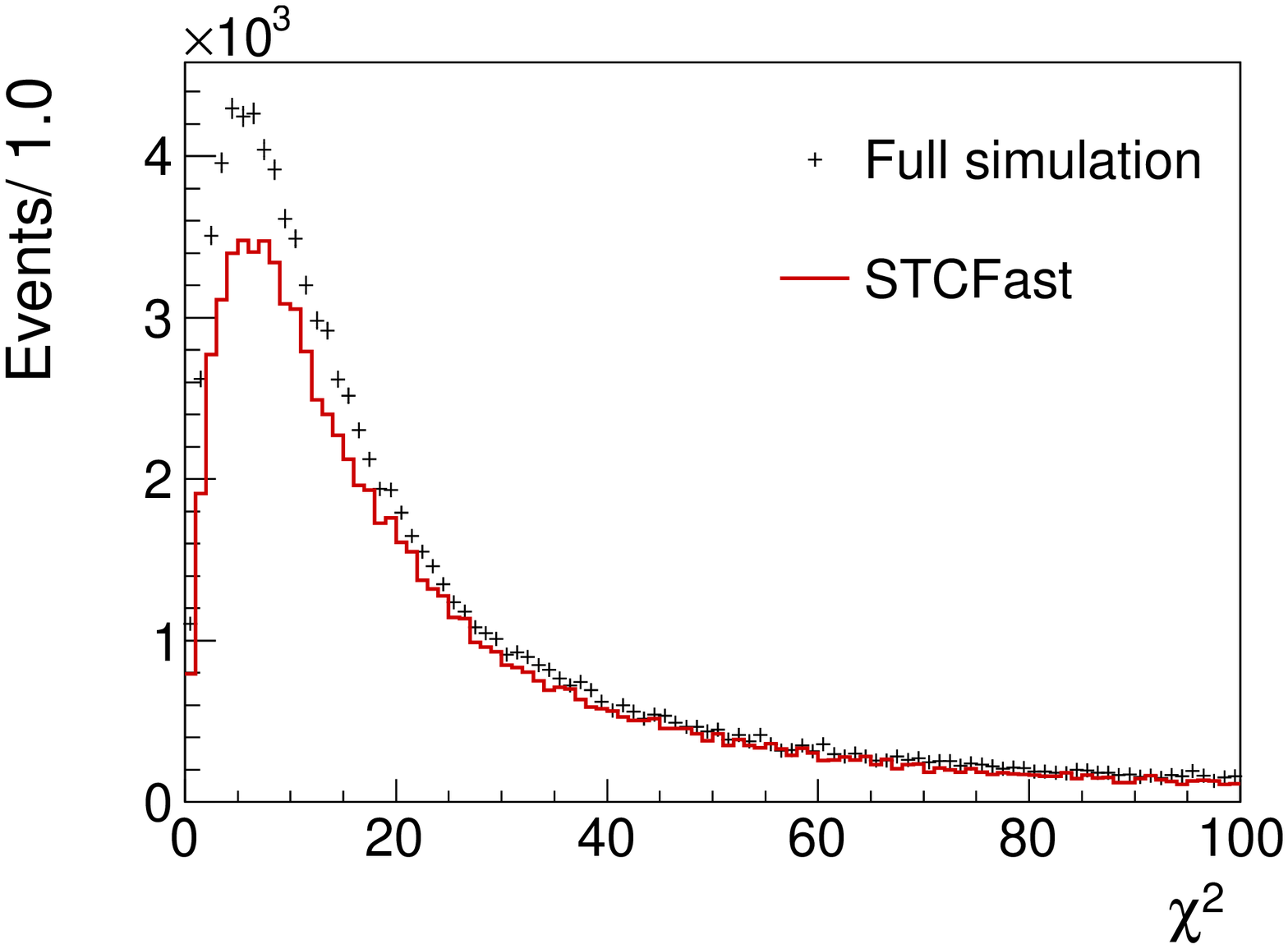}
\put(26,58){\small{(b)}}
\end{overpic}
\end{center}
\vskip -0.6cm
\caption{    (a) The distribution of $L/\sigma_{L}$ of $K_{S}^{0}$ from secondary vertex fit and (b) the $\chi^{2}$ distribution
from kinematic fit.
                Dots with error bars represent full simulation.
                The red line represents fast simulation.}
    \label{fit}
\end{figure}

The validation of long-lived particles ($K_{S}^{0}$, $\Lambda$)
is studied by comparing the discriminator observable, the flight length significance 
$L/\sigma_{L}$, where $L(\sigma_{L})$ is the flight length (uncertainty) reconstructed by a secondary vertex fit.
Fig.~\ref{fit}~(a) shows the comparison of $L/\sigma_{L}$ for $K_{S}^0$ between fast simulation and full simulation.

The kinematic fit is validated by the process $J/\psi\to\pi^{+}\pi^{-}\pi^{+}\pi^{-}$.
The comparison of the $\chi^{2}$ distribution of the kinematic fit is shown in
Fig.~\ref{fit}~(b) between the fast simulation and full simulation.
Since the correlation between the non-diagonal parameters in the helix error matrix of charged particles
is not considered in the fast simulation, as a consequence, the $\chi^{2}$ distribution for fast simulation is wider than that of the full simulation.
The discrepancy on $\chi^2$ depends on the
multiplicity of charged tracks in the kinematic fit: 
more tracks results in a larger discrepancy.

%%%%%%%%%%%%%
\section{Preliminary Results for STCF Detector}
%%%%%%%%%%%%%%%

As mentioned before, one major advantage of this fast simulation package is that the response of each sub-detector can be changed flexibly.
Thus, the package is very useful for the optimization of detector design during R\&D.
A few features are introduced below.

%%%%%%%%%%%%%%%%%
\subsection{ $\pi^{0}$ reconstruction}
\begin{figure}[htbp]
\begin{center}
\begin{overpic}[width=4.2cm,angle=0]{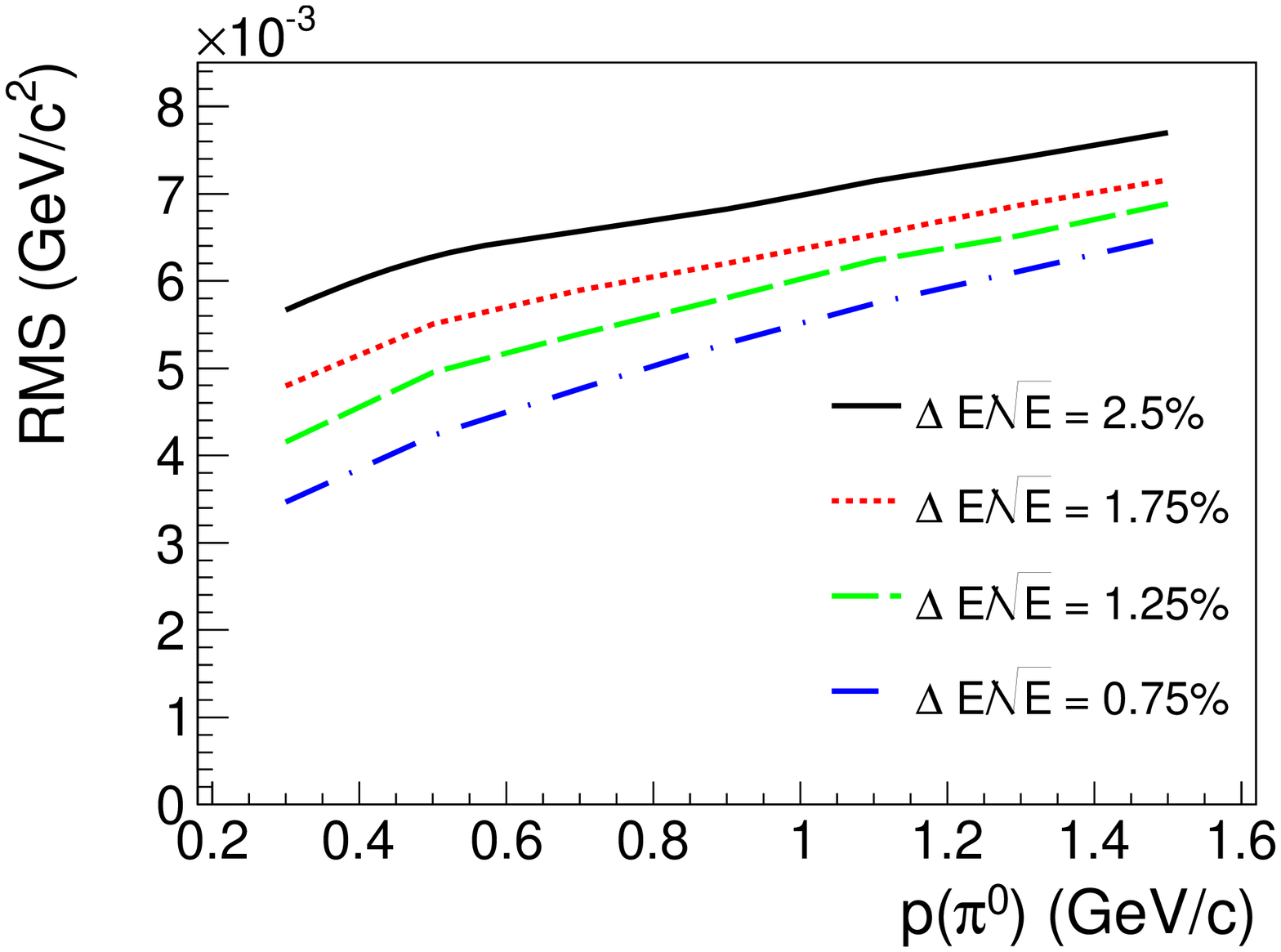}
\put(26,58){\small{(a)}}
\end{overpic}
\begin{overpic}[width=4.2cm,angle=0]{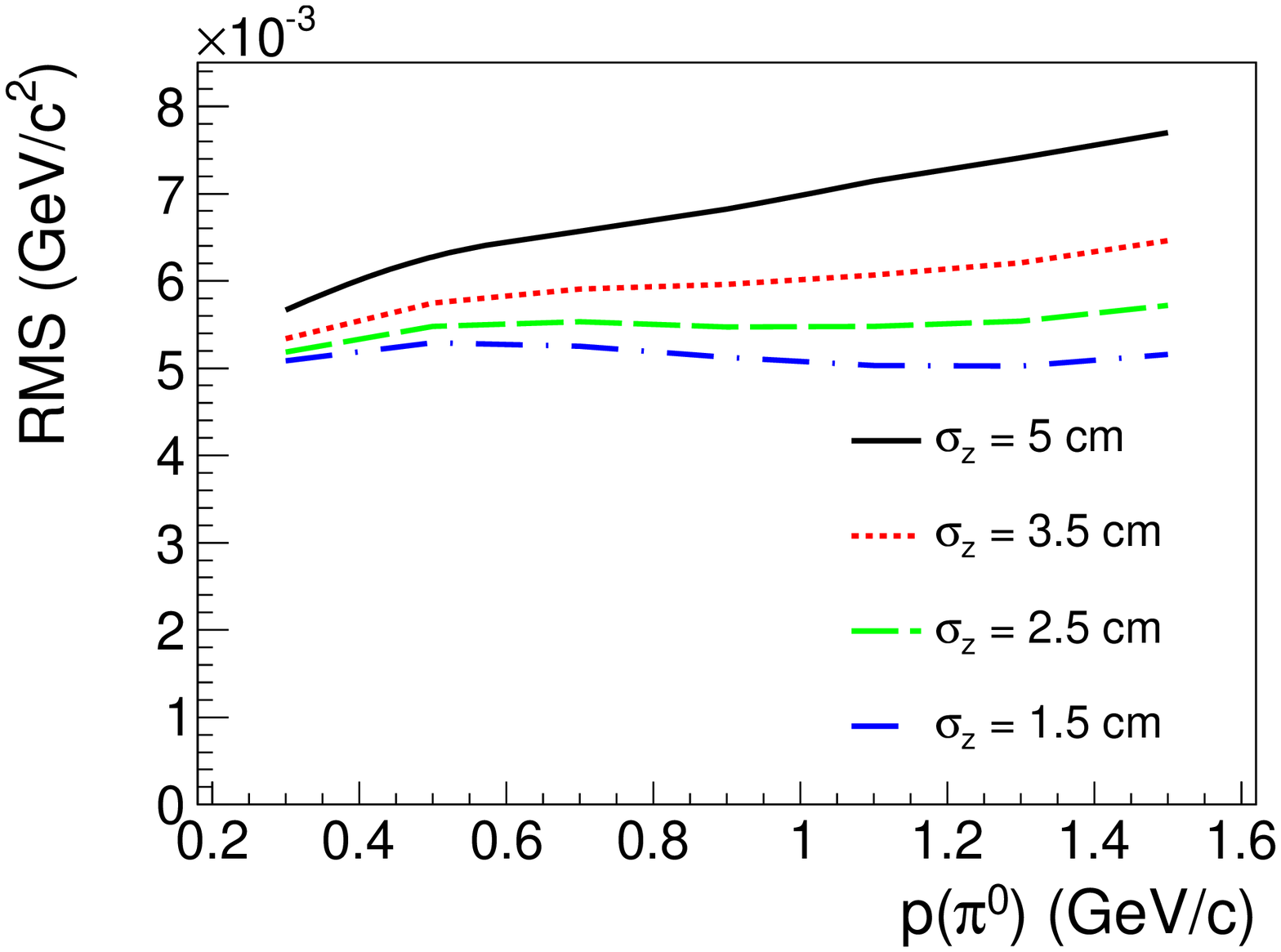}
\put(26,58){\small{(b)}}
\end{overpic}
\end{center}
\vskip -0.6cm
\caption{ The RMS of $M_{\pi^0}$ changed with (a) energy resolution at 1~$\gevc$
		and with (b) spatial resolution.
		The different color lines represent different resolutions.}
    \label{mpi0}
\end{figure}

%%%%%%%%%%%%%
There are many physical processes containing $\pi^{0}$s,
therefore it is important to improve the resolution of $\pi^{0}$s
which can have better signal to background ratio and higher detection efficiency in return.
Most $\pi^{0}$s can be reconstructed by two photons.
Using the interface to change the response of energy/position resolution of photon in the fast simulation package,
the evolutions of $M_{\pi^0}$ are studied along with these responses in its different momentum region.
The root mean square (RMS) value of $M_{\pi^0}$ show that for low momentum $\pi^0$s, the invariant mass resolution will be significantly improved with
higher energy resolution of photon,
while for high momentum $\pi^0$ the mass resolution of $\pi^{0}$ will be significantly improved with
higher position resolution of photon as shown in Figs.~\ref{mpi0}.

%%%%%%%%%%%%%%%%%
\subsection{Charged track reconstruction}
\begin{figure}[htbp]
\begin{center}
\begin{overpic}[width=4.2cm,angle=0]{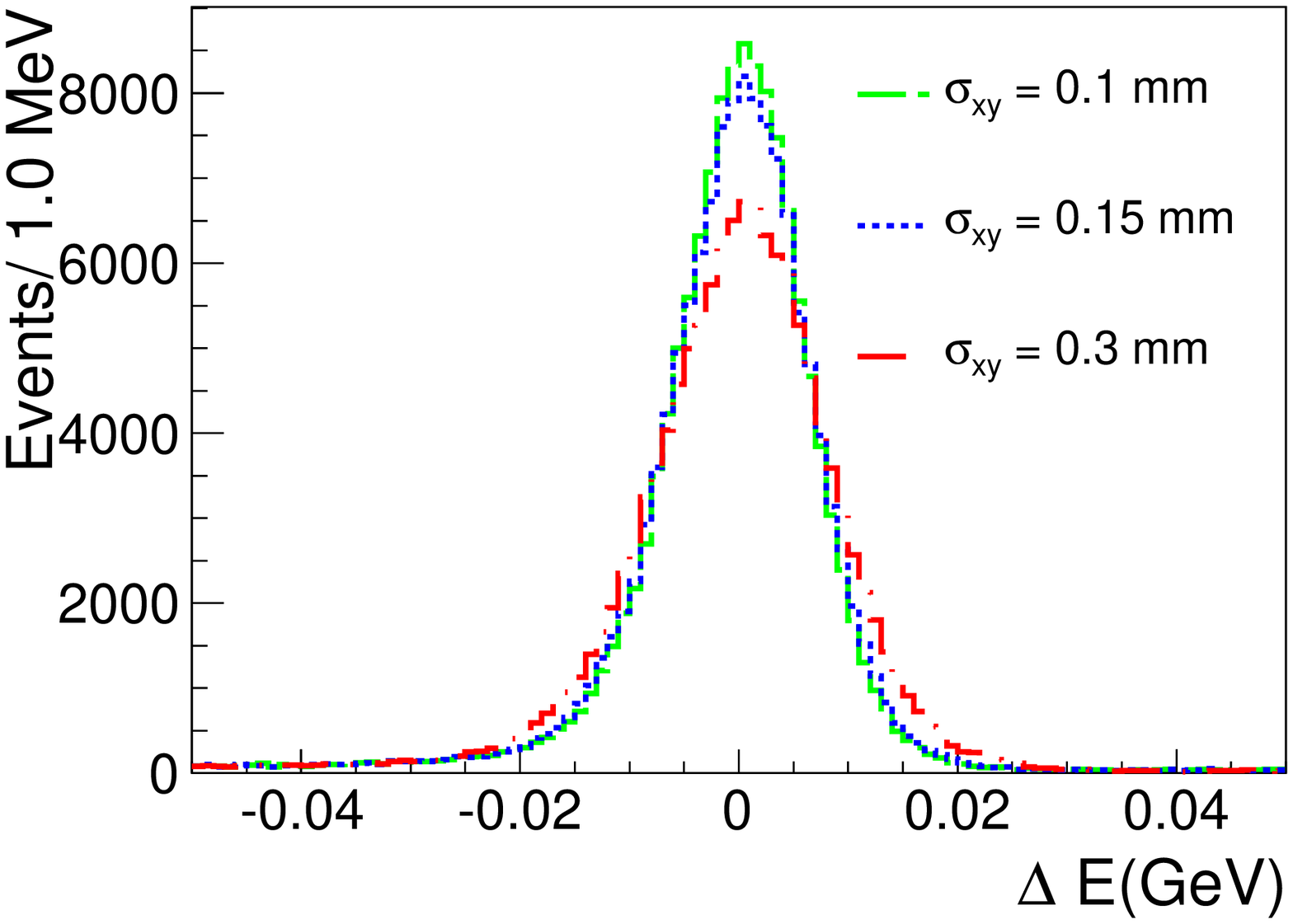}
\put(30,58){\small{(a)}}
\end{overpic}
\begin{overpic}[width=4.2cm,angle=0]{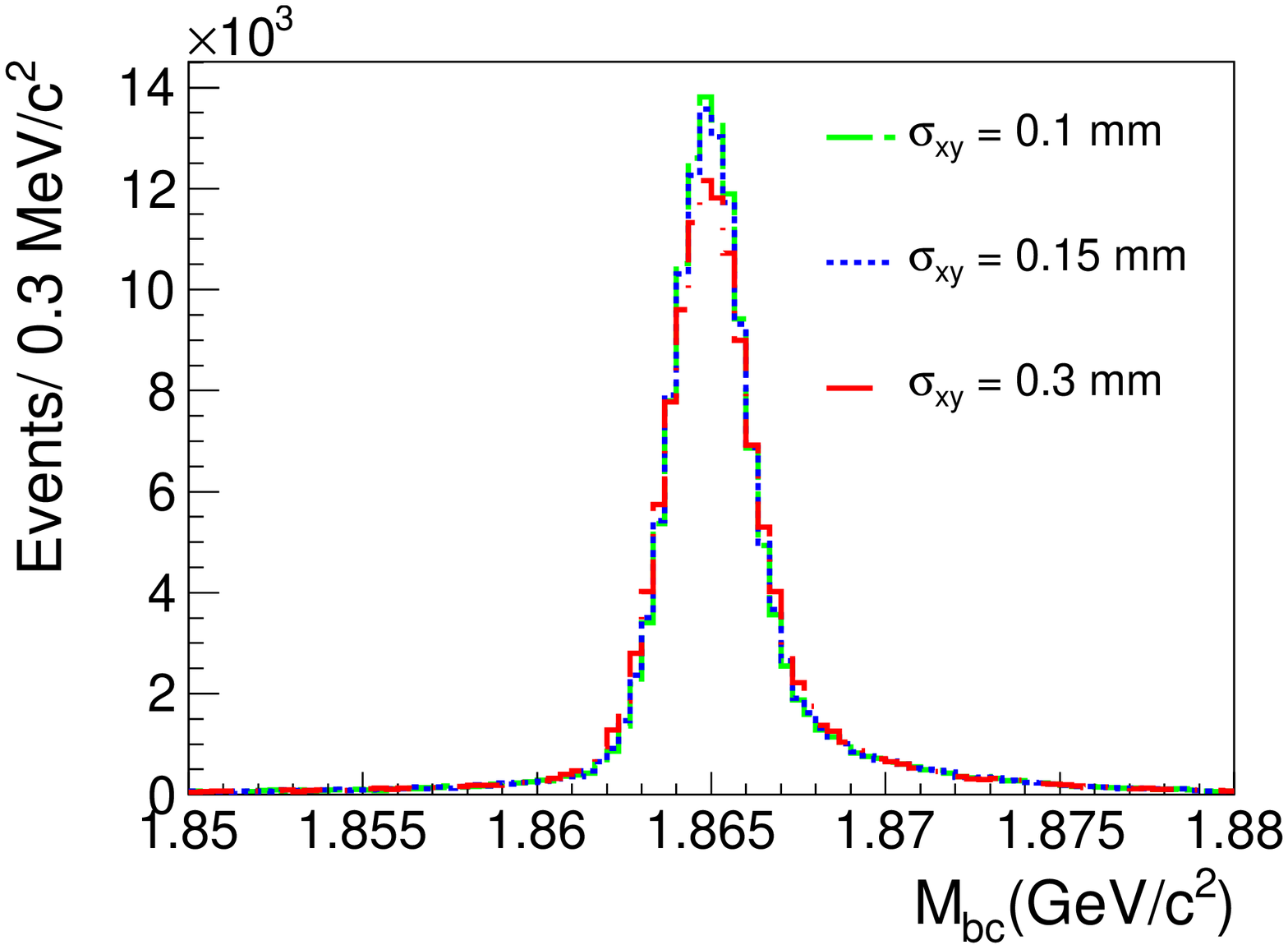}
\put(30,58){\small{(b)}}
\end{overpic}
\end{center}
\vskip -0.6cm
\caption{ 
	The distribution of (a) $\Delta E$ and (b) $M_{BC}$ dependence on the spatial resolution of the track system, 
	for the process  $e^{+}e^{-}\to D^{0}\bar{D}^{0}$ at $\sqrt{s}=3.77$~GeV with $D^0\to K^-\pi^+$.
		The different color lines represent different spatial resolutions.}
    \label{D0}
\end{figure}

%%%%%%%%%%%%%
At STCF, huge samples of charmed meson samples can be produced to make precision measurements of
various physical parameters such as measurement of the CKM matrix, $CP$ violation, and strong phase difference measurement, with unprecedented sensitivity.
In comparison with B factories, charm mesons like $D/D_{s}$ at STCF can be pair-produced near the production threshold, 
which presents a unique advantage of low backgrounds and high efficiency. 
Process production near threshold allows to exploit two key variables that
discriminate the signal from background, the energy difference $\Delta E=E-E_{beam}$ and the beam constrained mass $M_{BC} = \sqrt{E_{beam}^2/c^4-p^2/c^2}$,
where $E_{beam}$ is the beam energy, $E$ and $p$ are the total measured energy and three-momentum of the tagged charm meson, respectively.
The process $e^{+}e^{-}\to D^{0}\bar{D}^{0}$ at $\sqrt{s}=3.77$~GeV with $D^{0}$ reconstructed by $D^{0}\to K^{-}\pi^{+}$ is used 
to study the sensitivity of $\Delta E$ and $M_{BC}$'s resolution in association with the resolution of the track system~\cite{Drasal:2018zij}, 
The results indicate that a better spatial resolution of the charged track 
will improve the resolution of $\Delta E$ and $M_{BC}$, shown in Fig.~\ref{D0}.

%%%%%%%%%%%%%
\section{Conclusion and prospect}
%%%%%%%%%%%%%%%

In this paper, we introduce a fast simulation package dedicated to
study of the detector design and physical potential of the STCF.
%Comprehensive validation of the physics performance and detector response have done and
%good agreements between fast simulation and BESIII's full simulation are obtained.
By comparing the fast simulation and full simulation of the similar experiment BESIII, 
comprehensive validations are obtained.
The fast simulation provides a flexible approach for the user to change 
the response of each sub-detector system and has proven to be 
a useful tool for physics analysis.
Therefore, we recommend this fast simulation package for STCF phenomenological investigations~\cite{Shi:2019vus, Sang:2020}.

Several benchmark physics processes associated with key parameters of detector design are proposed for 
future study.
For example, the charged lepton flavor violation process via $\tau\to\gamma\mu$ or $\tau\to$ 3 leptons is sensitive
to the $\pi/\mu$ separation; the process to study the Collins fragmentation contribution
$e^{+}e^{-}\to h h'+X$, where $h/h'$ is $\pi$ or $K$ and $X$ denotes inclusive particles, is sensitive
to $\pi/K$ mis-identification~\cite{Ablikim:2015pta}; processes including $\pi^{0}$s are sensitive
to the EMC performance.
For the timing information of showers in EMC, processes including $n$ or $K_{L}^{0}$
could serve as benchmark processes.\\

%%%%%%%%%%%%%
%\section{Acknowledgment}
\section*{Acknowledgment}
The authors thank the supercomputing center of USTC and Hefei Comprehensive National Science Center for their strong support.
The authors also gratefully thank Prof.~Stephen Olsen for the advisement and proofreading.
The material presented in this paper is that of the authors alone, and has not been reviewed by the BESIII
collaboration; however, we thank our colleagues for allowing us to make use of the BESIII software environment.
The authors also thank the software group of BESIII and the detector group of STCF for the profitable discussions.
This work is supported by the Double First-Class university project foundation of USTC
and the National Natural Science Foundation of China under Projects No.11625523.

%\vspace{-1mm} \centerline{\rule{80mm}{0.1pt}} \vspace{2mm}

%\end{comment}


\begin{thebibliography}{90}

\vspace{3mm}

\bibitem{newphy}
  J.~D.~Lykken,
  %``Beyond the Standard Model,''
  CERN Yellow Report CERN-2010-002, 101-109
 % [arXiv:1005.1676 [hep-ph]].

\bibitem{confinement} J. Greensite, "An introduction to the confinement problem" Springer. ISBN 978-3-642-14381-6.
\bibitem{asymptotic1} D. J. Gross and F. Wilczek, Phys.\ Rev.\ Lett. {\bf 30}, 1343 (1973).
\bibitem{asymptotic2} H. D. Politzer Phys.\ Rev.\ Lett. {\bf 30}, 1346 (1973).

\bibitem{Adachi:2018qme}
  I.~Adachi {\it et al.} [Belle-II Collaboration],
  %``Detectors for extreme luminosity: Belle II,''
  Nucl.\ Instrum.\ Meth.\ A {\bf 907}, 46 (2018).
  %doi:10.1016/j.nima.2018.03.068

\bibitem{Alves:2008zz}
  A.~A.~Alves, Jr. {\it et al.} [LHCb Collaboration],
  %``The LHCb Detector at the LHC,''
  JINST {\bf 3}, S08005 (2008).
  %doi:10.1088/1748-0221/3/08/S08005


\bibitem{Galison:1992us}
  P.~Galison,
  %``Pure and hybrid detectors: Mark I and the psi,''
``The Rise of the standard model: Particle physics in the 1960s and 1970s``. Proceedings, Conference, Stanford, USA, June 24-27, 1992

\bibitem{Abrams:1989cm}
  G.~S.~Abrams {\it et al.},
  %``The Mark-{II} Detector for the {SLC},''
  Nucl.\ Instrum.\ Meth.\ A {\bf 281}, 55 (1989).
 % doi:10.1016/0168-9002(89)91217-5

\bibitem{Bernstein:1983wk}
  D.~Bernstein {\it et al.},
  %``The Mark-{III} Spectrometer,''
  Nucl.\ Instrum.\ Meth.\ A {\bf 226}, 301 (1984).
  %doi:10.1016/0168-9002(84)90043-3


\bibitem{Augustin:1980ad}
  J.~E.~Augustin {\it et al.},
  %``Dm2: A Magnetic Detector for the Orsay Storage Ring DCI,''
  Phys.\ Scripta {\bf 23}, 623 (1981).
  %doi:10.1088/0031-8949/23/4B/004

\bibitem{Bai:1994zm}
  J.~Z.~Bai {\it et al.} [BES Collaboration],
  %``The BES detector,''
  Nucl.\ Instrum.\ Meth.\ A {\bf 344}, 319 (1994).
  %doi:10.1016/0168-9002(94)90081-7
  %%CITATION = doi:10.1016/0168-9002(94)90081-7;%%
  %232 citations counted in INSPIRE as of 30 May 2019
\bibitem{Fang:2017znv}
  S.~Fang, G.~Huang and Z.~Zheng,
  %``$\tau$ mass and $R$-value measurements at BES,''
  Int.\ J.\ Mod.\ Phys.\ A {\bf 32}, no. 0, 1730004 (2017)
  %doi:10.1142/S0217751X17300046
  %[arXiv:1709.09545 [hep-ex]].
\bibitem{Ablikim:2005um}
  M.~Ablikim {\it et al.} [BES Collaboration],
  %``Observation of a resonance X(1835) in J / psi -> gamma pi+ pi- eta-prime,''
  Phys.\ Rev.\ Lett.\  {\bf 95}, 262001 (2005)
  %doi:10.1103/PhysRevLett.95.262001
  %[hep-ex/0508025].

\bibitem{bepcii}
C.\ Zhang for BEPC \& BEPCII Teams,
Performance of the BEPC and the progress of the BEPCII,
in: Proceeding of APAC, 2004, pp. 15-19, Gyeongju, Korea.

\bibitem{Ablikim:2009aa}
  M.\ Ablikim {\it et al.} [BESIII Collaboration],
	%``Design and Construction of the BESIII Detector,''
	\NIM \ A {\bf 614}, 345 (2010).

\bibitem{Ablikim:2019hff}
M.~Ablikim \textit{et al.} [BESIII],
%``Future Physics Programme of BESIII,''
Chin. Phys. C \textbf{44}, no.4, 040001 (2020)
%52 citations counted in INSPIRE as of 12 Oct 2020

\bibitem{Ablikim:2013mio}
  M.~Ablikim {\it et al.} [BESIII Collaboration],
  %``Observation of a Charged Charmoniumlike Structure in $e^+e^-$ → $π^+π^-$ J/ψ at $\sqrt{s}$ =4.26  GeV,''
  Phys.\ Rev.\ Lett.\  {\bf 110}, 252001 (2013)
  %doi:10.1103/PhysRevLett.110.252001
  %[arXiv:1303.5949 [hep-ex]].
\bibitem{BESIII:2012aa}
  M.~Ablikim {\it et al.} [BESIII Collaboration],
  %``First observation of $\eta(1405)$ decays into $f_{0}(980)\pi^0$,''
  Phys.\ Rev.\ Lett.\  {\bf 108}, 182001 (2012)
  %doi:10.1103/PhysRevLett.108.182001
  %[arXiv:1201.2737 [hep-ex]].
\bibitem{Ablikim:2016itz}
  M.~Ablikim {\it et al.} [BESIII Collaboration],
  %``Observation of an anomalous line shape of the $\eta^{\prime}\pi^{+}\pi^{-}$ mass spectrum near the $p\bar{p}$ mass threshold in $J/\psi\rightarrow\gamma\eta^{\prime}\pi^{+}\pi^{-}$,''
  Phys.\ Rev.\ Lett.\  {\bf 117}, no. 4, 042002 (2016)
 % doi:10.1103/PhysRevLett.117.042002
 % [arXiv:1603.09653 [hep-ex]].
\bibitem{Ablikim:2016sqt}
  M.~Ablikim {\it et al.} [BESIII Collaboration],
  %``Improved measurement of the absolute branching fraction of $D^{+}\rightarrow \bar{K}^0 \mu ^{+}\nu _{\mu }$,''
  Eur.\ Phys.\ J.\ C {\bf 76}, no. 7, 369 (2016)
 % doi:10.1140/epjc/s10052-016-4198-2
 % [arXiv:1605.00068 [hep-ex]].
\bibitem{Ablikim:2015orh}
  M.~Ablikim {\it et al.} [BESIII Collaboration],
  %``Measurement of the $e^+ e^− \to \pi^+ \pi^−$ cross section between 600 and 900 MeV using initial state radiation,''
  Phys.\ Lett.\ B {\bf 753}, 629 (2016)
 % doi:10.1016/j.physletb.2015.11.043
 % [arXiv:1507.08188 [hep-ex]].
\bibitem{Ablikim:2015flg}
  M.~Ablikim {\it et al.} [BESIII Collaboration],
  %``Measurements of absolute hadronic branching fractions of $\Lambda_{c}^{+}$ baryon,''
  Phys.\ Rev.\ Lett.\  {\bf 116}, no. 5, 052001 (2016)
  %doi:10.1103/PhysRevLett.116.052001
  %[arXiv:1511.08380 [hep-ex]].
	%\cite{Ablikim:2020yif}
\bibitem{Ablikim:2020yif}
M.~Ablikim \textit{et al.} [BESIII],
%``Determination of strong-phase parameters in $D\rightarrow K^0_{S,L}\pi^+\pi^-$,''
Phys. Rev. Lett. \textbf{124}, no.24, 241802 (2020)
%doi:10.1103/PhysRevLett.124.241802
%[arXiv:2002.12791 [hep-ex]].

%\cite{Ablikim:2020hsk}
\bibitem{Ablikim:2020hsk}
M.~Ablikim \textit{et al.} [BESIII],
%``Observation of a near-threshold structure in the $K^+$ recoil-mass spectra in $e^+e^-\to K^+ (D_s^- D^{*0} + D^{*-}_s D^0)$,''
[arXiv:2011.07855 [hep-ex]].

\bibitem{Kou:2018nap}
%  E.~Kou {\it et al.} [Belle-II Collaboration],
  %``The Belle II Physics Book,''
%  arXiv:1808.10567 [hep-ex].
E.~Kou {\it et al.} [Belle-II Collaboration],
  %``The Belle II Physics Book,''
  PTEP {\bf 2019}, no. 12, 123C01 (2019)
  Erratum: [PTEP {\bf 2020}, no. 2, 029201 (2020)]

%\cite{Bediaga:2018lhg}
\bibitem{Bediaga:2018lhg}
  R.~Aaij {\it et al.} [LHCb Collaboration],
  %``Physics case for an LHCb Upgrade II - Opportunities in flavour physics, and beyond, in the HL-LHC era,''
  arXiv:1808.08865.

\bibitem{Akai:2018mbz}
K.~Akai \textit{et al.} [SuperKEKB],
%``SuperKEKB Collider,''
Nucl. Instrum. Meth. A \textbf{907} (2018), 188-199
%doi:10.1016/j.nima.2018.08.017
%[arXiv:1809.01958 [physics.acc-ph]].
%54 citations counted in INSPIRE as of 14 Jan 2021

\bibitem{STCF}
H.-p.~Peng, ``High Intensity Electron Positron Accelerator (HIEPA),
Super Tau Charm Facility~(STCF) in China'', talk at Charm2018,
Novosibirsk, Russia, May 21 - 25, 2018.

\bibitem{Luo:2019gri}
  Q.~Luo,
  %``Progress of Preliminary Work for the Accelerators of a 2-7GeV Super Tau Charm Facility at China,''
  doi:10.18429/JACoW-eeFACT2018-TUOBB03
%\cite{Ohmi:2017cwi}
\bibitem{Ohmi:2017cwi}
K.~Ohmi, N.~Kuroo, K.~Oide, D.~Zhou and F.~Zimmermann,
%``Coherent Beam-Beam Instability in Collisions with a Large Crossing Angle,''
Phys. Rev. Lett. \textbf{119}, no.13, 134801 (2017)
%doi:10.1103/PhysRevLett.119.134801
%24 citations counted in INSPIRE as of 24 Aug 2020

\bibitem{Lewis:2018ayu}
P.~M.~Lewis, I.~Jaegle, H.~Nakayama, A.~Aloisio, F.~Ameli, M.~Barrett, A.~Beaulieu, L.~Bosisio, P.~Branchini and T.~E.~Browder, \textit{et al.}
%``First Measurements of Beam Backgrounds at SuperKEKB,''
Nucl. Instrum. Meth. A \textbf{914} (2019), 69-144
%doi:10.1016/j.nima.2018.05.071
%[arXiv:1802.01366 [physics.ins-det]].

\bibitem{Dong:2015kba}
M.~Y.~Dong, Q.~L.~Xiu, L.~H.~Wu, Z.~Wu, Z.~H.~Qin, P.~Shen, F.~F.~An, X.~D.~Ju, Y.~Liu and K.~Zhu, \textit{et al.}
%``Aging effect in the BESIII drift chamber,''
Chin. Phys. C \textbf{40} (2016) no.1, 016001
%doi:10.1088/1674-1137/40/1/016001
%[arXiv:1504.04681 [physics.ins-det]].

\bibitem{Petric:2017psf}
  M.~Petrič, M.~Frank, F.~Gaede, S.~Lu, N.~Nikiforou and A.~Sailer,
  %``Detector simulations with DD4hep,''
  J.\ Phys.\ Conf.\ Ser.\  {\bf 898}, no. 4, 042015 (2017).
 % doi:10.1088/1742-6596/898/4/042015

\bibitem{BOSS}
W.~D.~Li, Y.~J.~Mao and Y.~F.~Wang,
%``The BES-III detector and offline software,''
Int. J. Mod. Phys. A \textbf{24S1}, 9-21 (2009)
%doi:10.1142/S0217751X09046424

\bibitem{Ablikim:2017oaf}
  M.~Ablikim {\it et al.} [BESIII Collaboration],
  %``Measurement of $e^{+}e^{-}\rightarrow \pi^{+}\pi^{-}\psi(3686)$ from 4.008 to 4.600~GeV and observation of a charged structure in the $\pi^{\pm}\psi(3686)$ mass spectrum,''
  Phys.\ Rev.\ D {\bf 96}, no. 3, 032004 (2017)
  Erratum: [Phys.\ Rev.\ D {\bf 99}, no. 1, 019903 (2019)]
 % doi:10.1103/PhysRevD.96.032004, 10.1103/PhysRevD.99.019903
 % [arXiv:1703.08787 [hep-ex]].

\bibitem{BESIII} M. Ablikim {\it et al.} [BESIII Collaboration], Nucl.\ Instrum.\ Meth.\ A {\bf 614}, 3 (2010).


\bibitem{KKMC} S. Jadach, B. F. L. Ward, and Z. Was, Phys.\ Rev.\ D {\bf 63}, 113009 (2001).

\bibitem{EVTGEN} R. G. Ping, Chin.\ Phys.\ C {\bf 32}, 599 (2008);
                D. J. Lange, Nucl.\ Instrum.\ Meth.\ A {\bf 462}, 152 (2001).

\bibitem{pipipsip} M.~Ablikim {\it et al.} [BESIII Collaboration], Phys.\ Rev.\ D {\bf 96}, 032004 (2017).

\bibitem{PDG} M.~Tanabashi {\it et al.} [Particle Data Group], Phys.\ Rev.\ D {\bf 98}, 030001 (2018).


\bibitem{gampi0pi0}
  M.~Ablikim {\it et al.} [BESIII Collaboration],
  %``Amplitude analysis of the π$^0$π$^0$ system produced in radiative J/ψ decays,''
  Phys.\ Rev.\ D {\bf 92}, no. 5, 052003 (2015)
  Erratum: [Phys.\ Rev.\ D {\bf 93}, no. 3, 039906 (2016)]
%  doi:10.1103/PhysRevD.92.052003, 10.1103/PhysRevD.93.039906
%  [arXiv:1506.00546 [hep-ex]].
  %%CITATION = doi:10.1103/PhysRevD.92.052003, 10.1103/PhysRevD.93.039906;%%
  %24 citations counted in INSPIRE as of 30 May 2019

\bibitem{KM}
M.~Kobayashi and T.~Maskawa,
  %``CP Violation in the Renormalizable Theory of Weak Interaction,''
  Prog.\ Theor.\ Phys.\  {\bf 49}, 652 (1973).

%\cite{Drasal:2018zij}
\bibitem{Drasal:2018zij}
Z.~Drasal and W.~Riegler,
%``An extension of the Gluckstern formulae for multiple scattering: Analytic expressions for track parameter resolution using optimum weights,''
Nucl. Instrum. Meth. A \textbf{910}, 127-132 (2018)
%doi:10.1016/j.nima.2018.08.078
%[arXiv:1805.12014 [physics.ins-det]].
%2 citations counted in INSPIRE as of 28 Sep 2020

%\cite{Shi:2019vus}
\bibitem{Shi:2019vus}
X.~D.~Shi, X.~W.~Kang, I.~Bigi, W.~P.~Wang and H.~P.~Peng,
%``Prospects for CP and P violation in $\Lambda_{c}^+$ decays at Super Tau Charm Facility,''
Phys. Rev. D \textbf{100}, no.11, 113002 (2019)
%doi:10.1103/PhysRevD.100.113002
%[arXiv:1904.12415 [hep-ph]].
%3 citations counted in INSPIRE as of 28 Sep 2020

\bibitem{Sang:2020}
H.~Sang, X.~Shi, X.~Zhou, X.~Kang and J.~Liu,
%``Observation of a near-threshold structure in the $K^+$ recoil-mass spectra in $e^+e^-\to K^+ (D_s^- D^{*0} + D^{*-}_s D^0)$,''
[arXiv:2012.06241 [hep-ex]].

\bibitem{Ablikim:2015pta}
M.~Ablikim \textit{et al.} [BESIII],
%``Measurement of azimuthal asymmetries in inclusive charged dipion production in $e^+e^-$ annihilations at $\sqrt{s}$ = 3.65 GeV,''
Phys. Rev. Lett. \textbf{116} (2016) no.4, 042001
%doi:10.1103/PhysRevLett.116.042001
%[arXiv:1507.06824 [hep-ex]].

\end{thebibliography}
\end{document}